\documentstyle[sprocl]{article}

\newcommand{\eqn}[1]{(\ref{#1})}
\newcommand{\ft}[2]{{\textstyle\frac{#1}{#2}}}
\def\be{\begin{equation}}
\def\ee{\end{equation}}
\def\bea{\begin{eqnarray}}
\def\eea{\end{eqnarray}}
\renewcommand{\a}{\alpha}
\renewcommand{\b}{\beta}

\renewcommand{\d}{\delta}
\newcommand{\pa}{\partial}

\newcommand{\m}{\mu}
\newcommand{\n}{\nu}

\newcommand{\s}{\sigma}

\begin{document}
\begin{flushright}{THU-97/40\\NIKHEF~97-050\\[1mm] {\tt hep-th/9712082}}
\end{flushright}
\vspace{6mm}
\title{SUPERMEMBRANES AND SUPERMATRIX MODELS\footnote{Talk presented at
the Val\`encia workshop {\it Beyond the Standard Model; from
  Theory to Experiment}, October 13--17, 1997.}}

\author{B. DE WIT}

\address{Institute for Theoretical Physics,
Utrecht University,\\
Princetonplein 5, 3508 TA Utrecht, The Netherlands }

\author{K. PEETERS, J.C. PLEFKA}

\address{NIKHEF, P.O. Box 41882,\\
1009 DB Amsterdam, The Netherlands}


\maketitle\abstracts{We briefly review recent developments in the theory of 
supermembranes and supermatrix models. 
In a second part we discuss their interaction with  background fields. In
particular, we present the full background field coupling for the
bosonic case. This is a
short summary of the talk at the workshop. A more extended version
will appear elsewhere.}

\section{Supermembranes and matrix models}
It has been known for some time~\cite{DWHN} that certain supersymmetric
quantum-mechanical models characterized by the presence of zero-potential
valleys and Gauss-type constraints~\cite{CH} play an important
role in the quantization of fundamental supermembranes \cite{BST}.
In the light-cone formulation with a flat target 
space~\cite{BSTa,DWHN}, the supermembrane theory 
exhibits a residual invariance under area-preserving 
diffeomorphisms of the membrane surface. This
infinite-dimensional group can be truncated in many cases to a
finite-dimensional group, e.g. to U($N$), such as to lead to a
quantum-mechanical model based on a finite number of degrees of
freedom. The relevant Hamiltonian equals
\begin{equation} \label{hamiltonian}
H = g^{-1}\,{\rm Tr}\Big[ \ft{1}{2}{\bf P}^2 -
\ft{1}{4}[X^a,X^b]^2
+ \ft{1}{2} g\,\theta^{\rm T} \gamma_a [ X^a, \theta ]\Big] \, .
\end{equation}
Here, $\bf X$, $\bf P$ and $\theta$ take values in the Lie
algebra of the gauge group and are represented by matrices that
span this Lie algebra. Furthermore, these momenta and coordinates 
transform as vectors and spinors under the `transverse' SO(9) rotation 
group.\footnote{%
  The supermembrane can live in $D=11, 7, 5$ or 4 space-time
  dimensions, so that the transverse rotations are SO($D-2$). Here
  we restrict our attention exclusively to $D=11$.} %
The Hamiltonian \eqn{hamiltonian} can also  be interpreted as the
zero-volume limit of 10-dimensional supersymmetric Yang-Mills
theory with a corresponding gauge group. As explained above,
for the supermembrane this gauge group  consists of
area-preserving diffeomorphisms. The coupling constant $g$ is
then equal to the light-cone momentum of 
the membrane, denoted by $P_0^+=(P_-)_0$. 
 
Physical states must be invariant under the gauge group. Classical 
zero-energy configurations require all the commutators
$[X^a,X^b]$ to vanish. Dividing out the gauge group implies that 
zero-energy configurations are 
parametrized by ${\bf R}^9/S_N$. For the membrane, zero-energy
configurations correspond to zero-area stringlike configurations
of arbitrary length~\cite{DWLN}.  

At the quantum level the models described by 
\eqn{hamiltonian} have a continuous energy 
spectrum~\cite{DWLN,Smilga}. Whether or not a normalizable ground
state exists at the beginning of the continuum is a subtle
issue\footnote{%
  For discussions on the existence of massless states, see
  refs.\ 1, 7--9. According to ref.\ 9
  such states do indeed exist in $D=11$. }. %
Such a ground state must be annihilated by the supercharges. For
the supermembrane, it is expected to comprise the physical states 
of 11-dimensional supergravity. 

More recently it was shown that the very same quantum-mechanical
matrix models based on U$(N)$ describe the short-distance
dynamics of $N$ D0-branes~\cite{boundst}. Further interest in
these models has been triggered by a
conjecture according to which the degrees of freedom captured in M-theory,
are in fact described by the U$(N)$ supersymmetric matrix and 
related models in the $N\to \infty$ limit~\cite{BFSS}. 
M-theory is defined as the strong-coupling limit of type-IIA
string theory and is supposed to contain all the relevant degrees
of freedom of the known string theories, both at the perturbative
and the
nonperturbative level~\cite{Townsend,witten3,horvw,Mtheory}. In
this description the various string-string dualities play a
central role. At large distances M-theory is described by
11-dimensional supergravity~\cite{CJS}. 

So it turns out that M-theory, supermembranes and super-matrix
models are intricately related. A direct relation between
supermembranes and type-IIA theory was emphasized in particular
by Townsend~\cite{Townsend}, based on the relation between $d=10$ extremal
black holes in 10-dimensional supergravity and the Kaluza-Klein
states of 11-dimensional supergravity. {}From the string
point of view these states carry Ramond-Ramond charges, just as
the D0-branes~\cite{Polchinski}. Strings can
arise {}from membranes by a so-called double-dimensional
reduction~\cite{DHIS}. Supermembranes were, for example, also employed
to provide evidence for the duality of M-theory on $\mbox{\bf
R}^{10}\times S^1/\mbox{\bf Z}_2$ and 10-dimensional ${\rm E}_8\times
{\rm E}_8$ heterotic strings~\cite{horvw}.

An obvious question is whether fundamental supermembranes
themselves can provide the degrees of freedom of M-theory. This
question is often answered negatively in view of a suspected
quantum-mechanical inconsistency of the supermembrane theory.
Here we note that not too much is known about the 
actual consistency or
inconsistency of the supermembrane theory, which, incidentally, is linked
directly~\cite{GoldstoneHoppe} to the large-$N$ behaviour of the U($N$) matrix
models.  A possibly worrisome feature is also, that the fundamental
supermembrane does not seem to leave much room for its dual (`magnetic')
partner, the fivebrane. Irrespective on how one judges these
potential shortcomings, supermembranes give rise to an independent
perspective on many aspects of M-theory. Furthermore they have
many features in common with matrix models.

{}From a more technical point of view, the supermembrane provides us
with a well-defined tool for tackling various questions. For instance
the spherical supermembrane 
captures the nature of the $N\to \infty$ ``decompactification''
limit for the matrix models as well as the emergence of the
11-dimensional super-Poincar\'e invariance~\cite{DWMN,EMM}. One can also study
open membranes~\cite{openmb,Astrings} and topologically 
nontrivial membranes with or without  
winding around compact directions~\cite{us,Astrings}. As it turns
out, it is not always guaranteed that one can obtain a
matrix-model prescription in these situations. For instance, the
approach followed for 
compactification of matrix models, which is strongly guided by the
application to D-branes and T-duality \cite{Tdual}, does
not make immediate contact with winding supermembranes,
albeit that there are certain similarities. At present 
the significance of this situation is not completely clear to us.

A rather prominent question concerns matrix
models in curved backgrounds with a target-space tensor 
field~\cite{oldBG}.  
In the second part of this talk we demonstrate how
such couplings can be obtained {}from supermembrane theory 
formulated in a curved superspace background.

\section{Supermembranes in curved backgrounds}  
Let us consider a generic superspace with coordinates $Z^M =
(X^\m, \theta)$. World indices are $M=(\m,\a)$ and corresponding
tangent-space indices are denoted by $A=(r,a)$. The supermembrane
requires super-vielbeine $E_M^{\;A}$ and a 3-rank tensor gauge
field $B_{MNP}$. 
The action for a supermembrane~\cite{BST} is written in terms of embedding
coordinates $Z^M(\zeta)$, which are functions of the three
world-volume coordinates $\zeta^i$. It takes the following form,
\be
S=\int {\rm d}^3\zeta \;\Big[- \sqrt{-g(Z(\zeta))} - \ft16
\varepsilon^{ijk}\, \Pi_i^A \Pi_j^B\Pi_k^C \,B_{CBA}(Z(\zeta)) \Big]\,,
\label{supermem}
\ee
where $\Pi^A_i = \pa Z^M/\pa\zeta^i \; E_M^{\;A}$ and the induced
metric equals $g_{ij} =\Pi^r_i \Pi^s_j \,\eta_{rs}$, with
$\eta_{rs}$ the constant Lorentz-invariant metric. 

Flat superspace is characterized by
\renewcommand{\arraystretch}{1.4}
\begin{equation}
\begin{array}{rclcrcl}
E_\mu{}^r       &=& \d_\mu{}^r \, ,           
                & & E_\mu{}^a &=&0  \, ,\nonumber \\
E_\alpha{}^a    &=& \delta_\alpha{}^a  \, ,   
                & & E_\alpha{}^r &=& -(\bar\theta \Gamma^r)_{\alpha} \, ,\nonumber \\
B_{\mu\n\alpha} &=& (\bar\theta\Gamma_{\m\n})_{\alpha}\, , 
                & & B_{\mu\alpha\beta} &=& (\bar\theta\Gamma_{\mu\nu}){}_{(\alpha}
                                           (\bar\theta\Gamma^\nu){}_{\beta)}\, ,\\
B_{\a\b\gamma}  &=& (\bar\theta\Gamma_{\mu\nu})_{(\alpha} 
                                       (\bar\theta\Gamma^\mu)_{\vphantom{(}\beta} 
                                       (\bar\theta\Gamma^\nu)_{\gamma)}\, ,
                & & B_{\mu\nu\rho} &=& 0  \, .
\end{array}
\end{equation}
\renewcommand{\arraystretch}{1}
These quantities receive corrections in 
the presence of background fields. Here we consider a background 
induced by a nontrivial target-space metric, a target-space 
tensor field and a target-space gravitino field, corresponding to 
the fields of (on-shell) 11-dimensional supergravity. This
background can in principle be incorporated into superspace by a
procedure known as `gauge completion'~\cite{gc}. For 11-dimensional 
supergravity, the first steps of this procedure have been carried 
out long ago~\cite{CF}, but unfortunately only to first order in fermionic 
coordinates $\theta$. 
Meanwhile we have also determined the second-order contributions, 
which will be reported elsewhere.

To elucidate our general strategy, let us 
just confine ourselves to the purely bosonic case and present
the light-cone formulation of the membrane in a 
background consisting of the metric $G_{\mu\nu}$ and the tensor 
gauge field $C_{\mu\nu\rho}$.
The Lagrangian density for the bosonic membrane follows directly 
{}from \eqn{supermem},  
\begin{equation}
{\cal L} = -\sqrt{-g} + \ft{1}{6}\varepsilon^{ijk} \partial_i X^\mu\,
\partial_j
X^\nu \,\partial_k X^\rho\, C_{\mu\nu\rho} \, ,
\end{equation}
where $g_{ij}= \pa_i X^\m \,\pa_j X^\n \,\eta_{\m\n}$.  
For the light-cone formulation, the coordinates are decomposed in 
the usual fashion as $(X^+,X^-,X^a)$ with  
$a=1\ldots 9$. Furthermore we use the diffeomorphisms in the 
target space to bring the metric in a convenient form~\cite{GS},
\begin{equation}
\label{metricgauge}
G_{--}=G_{a-}=0\, .
\end{equation}
Subsequently we identify the time coordinate of the target space 
with the world-volume time, by imposing the condition  $X^+ = 
\tau$. Following the same steps as for the membrane in flat 
space~\cite{DWHN}, one arrives at a Hamiltonian formulation of 
the theory in terms of coordinates and momenta. These
phase-space variables are subject 
to a constraint. It is a pleasant surprise that this constraint 
takes the same form as for the membrane theory in flat space, namely,  
\begin{equation}
\phi_r = P_a \partial_r X^a + P_- \partial_r X^- \approx 0\, .
\end{equation}
Of course, the definition of the momenta in terms of the 
coordinates and their derivatives does involve the 
background fields, but at the end all explicit dependence 
on the background cancels out. 

The Hamiltonian now follows straightforwardly. As it turns out,
the background tensor field appears in the combinations
\bea
C_a &=& - \varepsilon^{rs} \partial_r X^- \partial_s X^b \,C_{-ab} +
\ft{1}{2}
\varepsilon^{rs}\partial_r X^b \partial_s X^c \,C_{abc} \, ,      
\nonumber \\
C_{\pm} &=& \ft{1}{2}\varepsilon^{rs}\partial_r X^a \partial_s X^b
\,C_{\pm ab}\,, \nonumber\\
C_{+-} &=& \varepsilon^{rs}\partial_r X^- \partial_s X^a\,
C_{+-a}\,.
\eea
We can further simplify the background by imposing a 
gauge where
\begin{equation}
C_{+-a} = 0\, , \quad C_{-ab}=0\, ,\quad G_{+-}=1\, .
\end{equation}
In that case the Hamiltonian takes the form
\bea
H&=& \int {\rm d}^2\sigma\, \Big \{
\frac{1}{P_-}\Big[\ft{1}{2}(P_a-C_a-P_-\, G_{a+})^2+ \ft{1}{4}
(\varepsilon^{rs}\, \partial_r X^a\, \partial_s X^b )^2\Big]\nonumber\\
&&\hspace{13.5mm}  -\ft{1}{2} P_-\, G_{++}-  \ft{1}{2}\varepsilon^{rs}\, 
\partial_r X^a\, \partial_s X^b\, C_{+ab} \Big \}\,,
\eea
where the coordinates $\s^{1,2}$ parametrize the spacesheet of 
the membrane.

As we want to avoid explicit time dependence of the background
fields, we assume the metric and the tensor field to be independent
of $X^+$. If we assume, in addition, that they are independent of
$X^-$, it turns out that $P_-$ becomes $\tau$-independent. This
allows us 
to set $P_-(\s)=(P_-)_0\,\sqrt{w(\s)}$, exactly as in flat space. Here 
$\sqrt w$ is some density on the spacesheet, whose  
integral is normalized to unity. With these restrictions,
it is possible to write down a gauge theory of area-preserving 
diffeomorphisms for the membrane in the presence of background 
fields. Its Lagrangian density equals  
\bea
{w}^{-1/2}\, {\cal L} &=& \ft{1}{2} (D_0 X^a)^2 + D_0 X^a \left(
\ft{1}{2} C_{abc} \{ X^b, X^c \} + G_{a+} \right) \nonumber \\
&&- \ft{1}{4}\{ X^a, X^b \}^2 + \ft12{G_{++}} + \ft{1}{2}
C_{+ab} \{ X^a, X^b \} \, ,
\eea
where we used the metric $G_{ab}$ to contract transverse
indices. Furthermore we adopted the usual definitions for the
Poisson bracket and the covariant derivative, 
\bea
\{ A,B\} &=& \frac{ \varepsilon^{rs}}{\sqrt{w}}\,\partial_r A\, \partial_s
B \, ,\nonumber\\ 
D_0\, X^a&=& \partial_0\, X^a- \{ \omega, X^a\} \, .
\eea
For convenience we have set $(P_-)_0 =1$. The above Lagrangian density 
is manifestly invariant under area-preserving diffeomorphisms in 
the presence of the background fields. It 
is now straightforward to write it in terms of a matrix model, 
by truncating the mode expansion for coordinates and momenta 
in the standard fashion~\cite{GoldstoneHoppe,DWHN}. A more explicit
derivation of these  
results and their supersymmetric generalization will 
appear in a separate publication~\cite{DWPP}. 

While preparing this report, we received two papers dealing with matrix 
models in certain backgrounds~\cite{Doug}.  We believe that the background  
interactions proposed in these papers are contained in our above results.

\section*{Acknowledgements}
B. de Wit thanks the organizers for
inviting him to this stimulating workshop.

\end{document}